\newcommand{\mathrm}[1]{{\rm #1}}

\documentstyle[aps,preprint,eqsecnum]{revtex}

\begin{document}
\preprint{}


\title{
MANY-BODY EFFECTS IN $^{16}\mathrm{O}(e,e'p)$}

\author{M. Radici, S. Boffi}
\address{
Dipartimento di Fisica Nucleare e Teorica, Universit\`{a} di Pavia, and\\
Istituto Nazionale di Fisica Nucleare, Sezione di Pavia, Pavia, Italy}

\author{Steven C. Pieper}
\address{
Physics Division, Argonne National Laboratory, Argonne, IL 60439}

\author{V. R. Pandharipande}
\address{
 Department of Physics, University of Illinois at Urbana-Champaign\\
1110 West Green Street, Urbana, IL 61801-3080}
\date{\today}

\maketitle
\begin{abstract}
Effects of nucleon-nucleon correlations on exclusive $(e,e'p)$
reactions on closed-shell nuclei leading to single-hole states are studied
using $^{16}\mathrm{O}(e,e'p)^{15}\mathrm{N}$ ($6.32$ MeV, $3/2^-$) as an
example.  The quasi-hole wave function, calculated from the overlap of
translationally invariant many-body variational wave functions containing
realistic spatial, spin and isospin correlations, seems to describe the initial
state of the struck proton accurately inside the nucleus, however it is too
large at the surface.  The effect of short-range correlations on the final
state
is found to be largely cancelled by the increase in the transparency for the
struck proton.  It is estimated that the values of the spectroscopic factors
obtained with the DWIA may increase by a few percent due to correlation effects
in the final state.
\end{abstract}

\pacs{PACS Nos }

\section{INTRODUCTION}

In the past decade the NIKHEF group \cite{van der Steenhoven} has accurately
measured cross sections on closed-shell nuclei for exclusive $(e,e'p)$
reactions leading to low-energy states in the residual nucleus. Some of these
states, denoted by $\vert\Psi_h\rangle$, can be regarded as having a
quasi-hole in a shell model orbital $h$.  The cross sections are in principle
written in terms of the transition matrix element of the nuclear charge-current
density operator $\hat{J}^\mu$ between the initial state $\vert\Psi_o\rangle$
describing the target nucleus and the final state $\vert\Psi_f\rangle$, which
asymptotically corresponds to an ejected proton leaving the residual nucleus in
the state $\vert\Psi_h\rangle$,

\begin{equation}
J^\mu = \int \mathrm{d}\vec{r} \> \mathrm{e}^{\mathrm{i} \vec{q} \cdot
\vec{r} } \langle \Psi_f \vert \hat{J}^\mu (\vec{r}\, ) \vert \Psi_o \rangle ,
\label{eq:ampl}
\end{equation}

\noindent
where $\vec{q}$ is the momentum transferred by the virtual photon.

Following the procedure of Ref.~\cite{BCCGP} it is possible to equivalently
rewrite the transition matrix in a one-body representation corresponding to
the specific channel selected by the experimental conditions. In momentum
space one has

\begin{equation}
J^\mu
= \int \mathrm{d}\vec{p} \  \psi_p^*(\vec{p}+\vec{q}\, )
\  \tilde J^\mu(\vec{p},\vec{q}\, ) \  \psi_h(\vec{p}\, )
\label{eq:eff}
\end{equation}

\noindent
where the advantage of working in a one-body representation is paid for by
the
introduction of an effective charge-current density operator $\tilde
J^\mu$. The single-particle wave function

\begin{equation}
\psi_h (\vec{p}\, ) = \langle \Psi_h \vert a (\vec{p}\, ) \vert
\Psi_o \rangle
\label{eq:specamp}
\end{equation}

\noindent
is the overlap between the target state $\vert \Psi_o \rangle$ and the hole
state $\vert \Psi_h \rangle$ produced by removing a particle with momentum
$\vec{p}$. It is characterized by the hole excitation energy $E_h$ and an
additional set of quantum numbers, which are here understood for simplicity.
In the following it will be called a quasi-hole wave function. The
normalization $S(E_h)$ of $\psi_h$ is the spectroscopic factor, which  measures
the probability that the state $\vert\Psi_h\rangle$ of the residual nucleus can
indeed be considered as a pure hole generated in the target nucleus by the
knockout process. A similar definition holds for the scattering state
$\psi_p$.

In principle, both wave functions are eigenfunctions of a Feshbach-like
non-local Hamiltonian referred to the residual nucleus \cite{BCCGP}. In
practice, calculations have been limited to reactions leading to low-lying
states which have a large overlap with single-hole states in the target
nucleus. The spectroscopic factors $S(E_h)$ for these states can often be
identified with the quasi-hole renormalization constant $Z$.  In such cases a
local energy-dependent mean-field complex potential $V(E,r)$ is adopted for
both
bound and scattering states \cite{BGP}. It is separately fitted to
single-particle bound state properties and to proton-nucleus
elastic-scattering
data. The non-locality of the Feshbach Hamiltonian is taken into account by
means of the Perey factor for both $\psi_h$ and $\psi_p$
\cite{Pereya,Pereyb,Capuzzi}

\begin{equation}
f_P (r) = \left( 1 - { {\partial V(E,r)} \over {\partial E} }
\right)^{1/2},  \label{eq:perey}
\end{equation}

\noindent
assuming a linear dependence of $V$ on the energy.
The quasi-hole $\psi_h$ of Eq. (\ref{eq:specamp}) is thus approximated as

\begin{equation}
\psi_h \sim \psi_{WS} = \sqrt{Z} f_P (r) \phi_{WS}(\vec{r}\, ) ,
\label{eq:ws}
\end{equation}

\noindent
where $\phi_{WS}(\vec{r}\, )$ is the wave function of the single particle state
$h$ at energy $E_h$ in a Woods-Saxon potential $V(E_h,r)$. The product
function $f_P (r) \phi_{WS}(\vec{r}\, )$ is normalized to unity so that $Z$ is
the normalization of $\psi_{WS}$.

The outgoing proton wave function is approximated as

\begin{equation}
\psi_p = f_P \chi
\label{eq:chiapp}
\end{equation}

\noindent
and is expanded in partial waves.
A Schr\"odinger equation including $V(E_p,r)$ is solved wave by wave for
$\chi$, where approximately $E_p \sim E_h + \omega$.
The observed cross sections are very well reproduced by
varying $Z$ and the parameters of the bound state Woods-Saxon potential
\cite{Lapikas}.

In this method of analysis the $(e,e'p)$ reaction is essentially considered as
a one-body process occurring in the average potential produced by the $(A-1)$
nucleons of the residual nucleus in the state $|\Psi_h\rangle$. Effects due
to two-body currents in $\hat{J}^\mu (\vec{p},\vec{q}\, )$ have been analyzed
in Refs.~\cite{BR,trieste}. If the mean-field
approximation were to be exactly valid, then the quasi-hole function would
have
normalization $Z = S(E_h) = 1$.  Typically $Z \sim 0.6$ is required to
reproduce the NIKHEF data on closed-shell nuclei from $^{16}\mathrm{O}$ to
$^{208}\mathrm{Pb}$
\cite{van der Steenhoven}.  This value of $Z$ is consistent with the observed
quenching of single-particle contributions to several other nuclear properties
\cite{Pandharipande}.

Correlations between nucleons in the nucleus reduce the value of $Z$, and
$1-Z$ can be regarded as a measure of their strength \cite{Pandharipande}.
The observed value of $Z$ suggests that correlations have a significant effect
on the $(e,e'p)$ reaction. Ideally, many-body calculations should then be
used to compute the $\psi_h, \psi_p$ and $\tilde J$ of Eq. (\ref{eq:eff}).

In the past decade, improved models of the nuclear Hamiltonian of the form

\begin{equation}
H = - {\hbar^2 \over 2m} \sum_{i} \nabla_i^2 + \sum_{i<j} v_{ij}
+ \sum_{i<j<k} V_{ijk}
\label{eq:hamilt}
\end{equation}

\noindent
have become available.  The two-nucleon interaction $v_{ij}$ is required
to reproduce the nucleon-nucleon scattering data, and the three-nucleon
interaction
$V_{ijk}$ is chosen to reproduce the binding energies of light nuclei and the
density of nuclear matter.  In the present work we use the Argonne
$\mathrm{v}_{14}$ model of $v_{ij}$ \cite{Wiringa} and the Urbana model VII of
$V_{ijk}$ \cite{Schiavilla}.

The ground state wave functions of $^3\mathrm{H}$ and $^4\mathrm{He}$
\cite{R.B.W.}, and $^{16}\mathrm{O}$
\cite{Pieper} have been obtained with this $H$ using the variational
method.  In the
few-body nuclei, the variational $|\Psi_o\rangle$ appears to be fairly
accurate from
comparisons with exact Green's function Monte Carlo calculations
\cite{Carlson},
and we can hope that they have useful accuracy for $^{16}\mathrm{O}$.
However, the
$|\Psi_f\rangle$ is much more difficult to calculate from a realistic $H$.
In order
to study $^4\mathrm{He}(e,e'p)^3\mathrm{H}$, measured at NIKHEF
\cite{van den Brand} and Saclay \cite{Magnon}, Schiavilla \cite{R.S.} used
the approximation

\begin{eqnarray}
|\Psi_f\rangle = &{\cal A} \{ {\cal S} {\displaystyle \prod_{i=1,3}}
{\cal F}_{4i} \} \, |\Psi_o(^3\mathrm{H} \ ,\ x_1,x_2,x_3)\  \chi(x_4)
\rangle \nonumber \\
  &+ \mbox{ {\rm orthogonality corrections.} } \label{eq:rocco}
\end{eqnarray}

\noindent
Here $x_i$ denotes $\vec{r}_i$,$\vec{\sigma}_i$ and $\vec{\tau}_i$ of the
$i$-th nucleon, ${\cal F}_{4i}$ denote correlations between the outgoing
nucleon and residual nucleons, and ${\cal A}$ and ${\cal S}$ are
antisymmetrization and symmetrization operators.  The $|\Psi_f\rangle$ is
orthogonalized to $|\Psi_o\rangle$ boosted with momentum transfer
$\vec{q}$.  The
$\chi(x_4)$ is determined from an optical potential, the correlation operator
${\cal F}_{4i}$ is estimated by using plane waves for particles 4 and $i$, and
the transition matrix element is calculated with the Monte Carlo method.

Such a calculation may be possible for $^{16}\mathrm{O}(e,e'p)$ reactions
leading
to the $p_{1/2}$ and
$p_{3/2}$ hole states of $^{15}\mathrm{N}$, for which cross sections have
been measured
by the NIKHEF group \cite{Leuschner}, using the available variational wave
functions $|\Psi_h\rangle$ for these $^{15}\mathrm{N}$ states
\cite{S.C.P.}.  However,
it is numerically much more difficult, and it will probably need an improved
treatment of
the correlation operator ${\cal F}_{Ai}$ describing correlations between
the struck,
high-energy nucleon $A$ and the bound nucleons $i=1,\ldots,15$ in
$^{15}\mathrm{N}$.

In the present work we consider a less ambitious treatment of this reaction
retaining a complete one-body charge-current operator \cite{BCCGP} for
$\tilde J$, using a many-body calculation for $\psi_h$ and considering some
approximate many-body effects on $\psi_p$.
Our two objectives are to test the
variational wave functions for $^{16}\mathrm{O}$ and $^{15}\mathrm{N}$ and
to estimate
the
effect
of plausible correlations in the final state on the analysis of
$(e,e'p)$ reactions.

The quasi-hole wave function, $\psi_h$, is calculated in Section II from the
overlap of the
variational wave functions of $^{16}\mathrm{O}$ and $^{15}\mathrm{N}$ using
methods
developed
earlier for helium liquid drops \cite{Lewart}. In the interior of the nucleus
this wave function is found to be very close to the $\psi_h$ obtained by
fitting
the NIKHEF data \cite{Leuschner}. However, it is too large in the surface
region,
suggesting that the variational wave functions of Refs. \cite{Pieper,S.C.P.}
do
not describe the nuclear surfaces very well.

In the traditional analysis of $(e,e'p)$, correlation effects on the initial
state of the struck proton are included through the $Z$ and by varying the
$\phi_{WS}$ to fit the data, but those on the final state are neglected.
In Section III the effects of
correlations in the final state are studied. We consider the following three
modifications of the final state:  (1) When the quasi-hole function $\psi_h$
is calculated from the overlap
of $|\Psi_o\rangle$ and $|\Psi_h\rangle$, the Perey factor $f_P (r)$
cannot be calculated from Eq. (\ref{eq:perey}).  It is identified with the
effective mass correction $\sqrt{m^*(r)/m}$ \cite{Mahaux}. (2) The $\chi$ is
multiplied by a factor $\sqrt{Z(\vec{r}\, )}$, estimated with the local density
approximation \cite{Lewart}, to take into account the effect of short-range
correlations between the struck proton and the residual nucleons.  (3) The
struck
proton is a part of the $^{16}\mathrm{O}$ ground state. Hence, its final state
interactions on the way out are driven by the ground state density weighted
by the pair distribution function. This ``correlation-hole'' effect is
observed in inclusive $(e,e'p)$ \cite{V.R.P.}
and $(e,e')$ \cite{Benhar} reactions, and it is estimated by multiplying the
$\chi(x)$ by a factor $\lambda(\vec{r}\, )$ calculated from the pair
distribution
function.

\section{THE QUASI-HOLE WAVE FUNCTION}

The variational wave function used in Ref.~\cite{Pieper} to describe the
ground state of $^{16}\mathrm{O}$ has the form

\begin{equation}
|\Psi_o\rangle = \prod_{IT} (1 + U_{ijk}) \{ S \prod_{i<j} (1 + U_{ij}\}
|\Psi_{J,o}\rangle,
\label{eq:varpsi}
\end{equation}

\begin{equation}
|\Psi_{J,o}\rangle = \prod_{i<j} f_c(r_{ij}) {\cal A} |\Phi_o\rangle.
\label{eq:jaspsi}
\end{equation}

\noindent
Here $|\Phi_o\rangle$ is an independent particle wave function, $f_c(r_{ij})$
represents spatial pair correlations, the operators $U_{ij}$ and $U_{ijk}$
describe
two- and three-particle spin, isospin, tensor and spin-orbit correlations,
$IT$
denotes a product of $(1 + U_{ijk})$ containing only independent triplets,
and $\cal S$ and $\cal A$ denote symmetrization and antisymmetrization
operators. The
$\Phi_o$ is a simple product of single-particle wave functions
$\phi_n(\vec{r}_i-\vec{R}_A)$ in the $A$-nucleon center of mass frame:

\begin{equation}
\vec{R}_A = {1 \over A} \sum_{i=1}^A \vec{r}_i.
\label{eq:cm}
\end{equation}

\noindent
It thus contains many-body correlations required to make $|\Phi_o\rangle$, and
therefore $|\Psi_o\rangle$, translationally invariant.

Wave functions for single-hole states of $^{15}\mathrm{N}$ are
approximately obtained
by
removing a nucleon from $\Phi_o$ in the state $h$.  The single-particle wave
functions of $|\Phi_h\rangle$ are $\phi_n(\vec{r}_i-\vec{R}_{A-1})$, where

\begin{equation}
\vec{R}_{A-1} = {1 \over (A-1)} \sum_{i=1}^{A-1} \vec{r}_i,
\label{eq:cmres}
\end{equation}

\noindent
so that $|\Phi_h\rangle$ is also translationally invariant.  The two- and
three-body
correlations in $^{15}\mathrm{N}$ hole states are assumed to be the same as in
$^{16}\mathrm{O}$ in
this approximation.  The $|\Psi_{J,h}\rangle$ and $|\Psi_h\rangle$ are
obtained from
$|\Phi_h\rangle$ with equations like (\ref{eq:varpsi}) and (\ref{eq:jaspsi}).

If the state that is removed from $\Phi_o$ has $\ell=1$, $m_{\ell}=-1$ and
$m_s=-1/2$, then the $|\Phi_h\rangle$ and $|\Psi_h\rangle$ have
$J,M=3/2,3/2$.  It
was argued in \cite{S.C.P.} that this state should be associated with the
centroid at
$\sim 6.87$ MeV of the discrete $3/2^-$ strength in $^{15}\mathrm{N}$.  It
was shown
that it
reproduces reasonably the $^{15}\mathrm{N}(3/2^-)-^{16}\mathrm{O}$ energy
difference
and,
with the
corresponding $1/2^-$ state, the spin-orbit splitting in $^{15}\mathrm{N}$.
 Here we
use
it to calculate the quasi-hole orbital.

The $\psi_h$ is function of $\vec{r} = \vec{r}_A-\vec{R}_{A-1}$ and can be
written as

\begin{equation}
\psi_h(x) = \psi_h(r){\cal Y}^m_{\ell sj}(\hat{r},\vec{\sigma}\, ) =
\langle\Psi_h|a(x)|\Psi_o\rangle
\label{eq:varqh}
\end{equation}

\begin{equation}
{\psi_h(r) \over 4\pi r^2} = \sqrt{A}
{\langle\Psi_h(x_1...x_{A-1}){\cal Y}^{m}_{\ell
sj}(x_A)\  \delta(r-|\vec{r}_A-\vec{R}_{A-1}|)\ |\Psi_o(x_1...x_A)\rangle
\over \langle\Psi_h|\Psi_h\rangle^{1/2}\  \langle\Psi_o|\Psi_o\rangle^{1/2}},
\label{eq:rvarqh}
\end{equation}

\noindent
where ${\cal Y}^m_{\ell sj}$ are standard spin-angle functions,
for example

\begin{equation}
{\cal Y}^{3/2}_{1 \mbox{ } 1/2 \mbox{ } 3/2} = Y_{11}(\hat{r})|m_s =
1/2\rangle.
\label{eq:spinang}
\end{equation}

\noindent
The factor $\sqrt{A}$ in Eq. (\ref{eq:rvarqh}) takes into account the
possibility that the removed particle can be any of the $A$ nucleons in
$|\Psi_o\rangle$.  The quasi-hole normalization $Z_h$ is given by

\begin{equation}
Z_h = \int r^2 dr \  \psi^2_h(r).
\label{eq:zeta}
\end{equation}

In momentum space the $\psi_h(p)$ is defined as

\begin{equation}
\psi_h(p) = \sqrt{A}
{\langle\Psi_h(x_1...x_{A-1})\  \eta(p,x_A)|\Psi_o(x_1...x_A)\rangle
\over \langle\Psi_o|\Psi_o\rangle} \sqrt{\langle\Psi_o|\Psi_o\rangle \over
\langle\Psi_h|\Psi_h\rangle},
\label{eq:pvarqh}
\end{equation}

\noindent
where

\begin{equation}
\eta(p,x_A) = j_{\ell}(p|\vec{r}_A-\vec{R}_{A-1}|) {\cal Y}^m_{\ell sj}.
\label{eq:eta}
\end{equation}

\noindent
The normalization of $\psi_h(p)$ is $Z_h\pi/2 \,$.

In Monte Carlo calculations both $\psi_h(r)$ and $\psi_h(p)$ are
simultaneously calculated in order to avoid Fourier transforms of data with
sampling errors.  For brevity we discuss the calculation of only
$\psi_h(p)$.  It is convenient to write Eq. (\ref{eq:pvarqh}) as

\begin{equation}
\psi_h(p) = \sqrt{A} \mbox{ } \mbox{ } Q(p)/\sqrt{M}
\label{eq:pvarqhmc}
\end{equation}

\noindent
and evaluate $Q$ and $M$ using cluster expansions \cite{Pieper}.  The
overlaps $\langle {\cal O} \rangle$ are defined as

\begin{equation}
\langle {\cal O} \rangle =
{\langle\Psi_{J,h}(x_1...x_{A-1})\ \eta(p,x_A)| {\cal O} |\Psi_{J.o}\rangle
\over \langle\Psi_{J,o}|\Psi_{J,o}\rangle}
\label{eq:meanval}
\end{equation}

\noindent
and calculated with the Monte Carlo method.  The cluster expansion of $Q$
is then given by

\begin{equation}
Q = q_o + \sum_{i<j} q_{ij} + \sum_{i<j<k} q_{ijk} +.... .
\label{eq:varq}
\end{equation}

\noindent
We obtain

\begin{equation}
q_o = \langle1\rangle,
\label{eq:qone}
\end{equation}

\begin{eqnarray}
q_{ij} &= {\displaystyle {\langle(1 + U^{\dag}_{ij})(1 + U_{ij})\rangle-q_o
\over
1 + d^A_{ij}} }
 \mbox{ }&\mbox{ for } j \neq A, \nonumber \\
 &= {\displaystyle {\langle U_{iA}\rangle \over 1 + d^A_{iA}} } \mbox{ }
\qquad
&\mbox{ for } j = A. \label{eq:qtwo}
\end{eqnarray}

\noindent
The $d^A_{ij}$ are given by

\begin{equation}
d^A_{ij} = \langle(1 + U^{\dag}_{ij})(1 + U_{ij})\rangle \, -1
\label{eq:dtwo}
\end{equation}

\noindent
for all $j$ and $i$.  For the sake of brevity we do not give detailed
expressions for $q_{ijk}$, $d^A_{ijk}$, etc.; they can be easily obtained from
methods given in Ref. \cite{Pieper}.

The ratio $M$ is written as

\begin{equation}
M = {\langle\Psi_h|\Psi_h\rangle \over \langle\Psi_o|\Psi_o\rangle} = M_uM_J,
\label{eq:varm}
\end{equation}

\noindent
and

\begin{equation}
M_u = {\langle\Psi_h|\Psi_h\rangle \over \langle\Psi_{J,h}|\Psi_{J,h}\rangle}
{\langle\Psi_{J,o}|\Psi_{J,o}\rangle \over \langle\Psi_o|\Psi_o\rangle}
\label{eq:varmu}
\end{equation}

\noindent
is evaluated using a cluster expansion.  Defining

\begin{equation}
d^{A-1}_{ij} = {\langle\Psi_{J,h}|(1 + U^{\dag}_{ij})(1 +
U_{ij})|\Psi_{J,h}\rangle \over \langle\Psi_{J,h}|\Psi_{J,h}\rangle} \, -1,
\label{eq:dtwores}
\end{equation}

\noindent
and $d^{A-1}_{ijk}$, etc., analogously we get

\begin{equation}
M_u =
 {\displaystyle 1 + \sum_{{i<j\leq A-1}}d^{A-1}_{ij} +
      \sum_{{i<j<k\leq A-1}}d^{A-1}_{ijk} +
      \ldots
      \over  \displaystyle
  1 + \sum_{{i<j\leq A}}d^A_{ij} +
      \sum_{{i<j<k\leq A}}d^A_{ijk} +
      \ldots}  .
\label{eq:varmubis}
\end{equation}

\noindent
In practice we express $M_u$ as a sum of irreducible cluster terms as in
Eq. (\ref{eq:varq}).  The ratio

\begin{equation}
M_J = {\langle\Psi_{J,h}|\Psi_{J,h}\rangle \over
\langle\Psi_{J,o}|\Psi_{J,o}\rangle}
\label{eq:varmjas}
\end{equation}

\noindent
can be calculated exactly without cluster expansions. It is written as

\begin{equation}
M_J =
{\langle\Psi_{J,h}\phi_h(\vec{r}_A-\vec{R}_{A-1})|\phi_h(\vec{r}_A-\vec{R}_{
A-1})\Psi_{J,h}\rangle
\over \langle\Psi_{J,o}|\Psi_{J,o}\rangle}
\label{eq:mjasbis}
\end{equation}

\noindent
for convenient Monte Carlo evaluation, using a normalized $\phi_h$:

\begin{equation}
\int |\phi_h(\vec{r}\, )|^2 \mathrm{d}\vec{r} = 1.
\label{eq:norm}
\end{equation}

The complete calculation of $\psi_h$ thus requires two separate Monte Carlo
walks:
(i) an $A$-body walk in which the $q$'s of Eq. (\ref{eq:varq}), the $d^A$'s of
Eq. (\ref{eq:dtwo}) and the $M_J$ of Eq. (\ref{eq:mjasbis}) are calculated;
and
(ii) an $(A-1)$-body walk in which the $d^{A-1}$'s of Eq. (\ref{eq:dtwores})
are calculated. Attempts to evaluate the $d^{A-1}$'s from the $A$-body walk
led
to larger sampling errors.

As discussed in Ref. \cite{Pieper}, the optimal variational $|\Psi_o\rangle$
that minimizes the ground state energy does not give a good representation
of the experimental $^{16}\mathrm{O}$ charge form factor.  It is possible
to reproduce
the observed $^{16}\mathrm{O}$ charge form factor by changing only the
one-body part
$|\Phi_o\rangle$ of
the optimum $|\Psi_o\rangle$.  In this paper we are interested in the study of
spatial wave functions with the $(e,e'p)$ reaction.  It therefore seems
reasonable to start with the wave function $|\Psi_o\rangle$ that gives an
accurate
description of the charge form factor and contains optimal correlation
operators
$f_c(r_{ij})$, $U_{ij}$ and $U_{ijk}$.  The results presented in this paper
are
obtained with such $|\Psi_o\rangle$ and $|\Psi_h\rangle$. However, none of our
conclusions change significantly when the optimal variational wave functions
are
used instead.  The $|\Psi_o\rangle$ constrained to reproduce the observed
charge form
factor of $^{16}\mathrm{O}$ gives $\sim 5\%$ less binding energy than the
optimal
wave function \cite{Pieper}.

The results for $\psi_h(p)$ are shown in Fig. 1.  The dotted line shows
$\phi^2_h(p)$, where $\phi_h$ is the single-particle wave function in
$\Phi_o$.  The norm of this wave
function is, of course, unity.  The dash-dot curve shows the
$\psi^2_{h,cm}(p)$
obtained when the center-of-mass effects are included, but no other
correlations.
In this approximation the $q_{ij}$, $d^A_{ij}$, $d^{A-1}_{ij}$, etc. are
all zero,
so that $M_u=M_J=1$  and

\begin{eqnarray}
Q(p) &= q_{o,cm}(p), \nonumber \\
\psi_{h,cm}(p) &= \sqrt{A}  \mbox{ } q_{o,cm}(p). \label{eq:qhcm}
\end{eqnarray}

\noindent
A full $A$-body integration is necessary to calculate the $\psi_{h,cm}$.
The $\psi^2_{h,cm}(p) < \phi^2_h(p)$  and its normalization is $Z_h=0.88$.
The dashed curve shows the
further effect of including central correlations $f_c(r_{ij})$ only.  In this
approximation

\begin{equation}
\psi_{h,J}(p) = \sqrt{A}  \mbox{ } q_o(p)/\sqrt{M_J},
\label{eq:qhjas}
\end{equation}

\noindent
with $\sqrt{M_J} = 1.013(3)$ and $Z_h=0.87$.
Fig. 1 shows that the function $\psi_{h,J}(p)$ is rather close to
$\psi_{h,cm}$.
Note that the present
calculation of $\psi_{h,cm}$ and $\psi_{h,J}$ from the chosen many-body wave
function is exact.

At present it is necessary to use cluster expansions to calculate overlaps
of wave
functions with spin-isospin correlations.  Two- and three-body cluster
contributions are retained in the calculation of the $\psi_h(p)$ from the
complete
variational wave functions $\Psi_o$ and $\Psi_h$.  The two-body terms
reduce $Z_h$
by $0.09$ to $0.78$, and the three-body terms increase $Z_h$ by $0.04$ to
its present
value of $0.82$.  We expect that a complete calculation will give $Z_h =
0.81\pm.02$
with present wave functions.  The $\psi^2_h(p)$ is shown by the full line in
Fig. 1.

M\"{u}ther and Dickhoff \cite{Muther} have recently studied the quasi-hole
orbitals in $^{16}\mathrm{O}$ using Brueckner theory.  They neglect the
effect of
center-of-mass motion and obtain $Z_h = 0.91$ for the  $p_{3/2}$ state.  We
obtain a
similar value $(Z_h = 0.90)$ by including $f_c(r_{ij})$, $U_{ij}$ and
$U_{ijk}$
correlations, but no center-of-mass corrections.  It is interesting that in
a light
nucleus like $^{16}\mathrm{O}$ the center-of-mass effects seem to account
for almost
half of the reduction of $Z_h$ from its unit value.

Most of the low-energy $p_{3/2}$ strength observed in $^{16}\mathrm{O}(e,e'p)$
reactions
goes to the $3/2^-$ states at $6.32$, $9.93$ and $10.7$ MeV in
$^{15}\mathrm{N}$
\cite{Leuschner}.  The state at $6.32$ MeV has $86\%$ of the total strength
in these
three states.  We assume that the $p_{3/2}$ quasi-hole state fragments due to
mixing
with other more complex states.  The quasi-hole wave function

\begin{equation}
\psi_{6.32} \equiv \langle \Psi(6.32\mbox{ MeV in}\
^{15}\mathrm{N})|a(x)|\Psi_o\rangle
\label{eq:qhexp}
\end{equation}

\noindent
is then given by

\begin{equation}
\psi_{6.32} = \sqrt{0.86}  \mbox{ }  \psi_h,
\label{eq:qhnorm}
\end{equation}

\noindent
assuming that the more complex states have negligible overlap with the state
$a(x)|\Psi_o\rangle$.  The empirical wave function $\psi_{WS}$,
obtained by fitting the
$(e,e'p)$
cross-section to the $6.32$ MeV state with the parameterization
(\ref{eq:ws}), is therefore
compared with $\sqrt {0.86}\,\psi_h$ in Fig.~2.  We note that at small $r$
the calculated and fitted wave functions are very similar, however, at
large $r$ the
$\psi_h$ is too large.  The $Z$ of $\psi_{WS}$ is $0.53$, whereas the
above many-body calculation gives

\begin{equation}
Z_{6.32} = 0.86 Z_h = 0.70.
\label{eq:normexp}
\end{equation}

These differences between the calculated and fitted wave functions are
indicative
of the limitations of the variational wave functions (\ref{eq:varpsi}).
They may not be able
to describe the clustering of nucleons in the surface of the nucleus.  Such a
clustering can also lead to fluctuations in the shape of the nucleus.

We attempted to study the possibility of surface vibrations reducing
the $Z_h$ by including a factor

\begin{equation}
\{ 1 + \sum_{\ell}  \mbox{ } \alpha_{\ell} \mbox{ }  \sum_{{i<j}}
    r_i^{\ell} r_j^{\ell} \mbox{ } \sum_{m,m'} \mbox{ } [ Y_{\ell m}(\hat{r_i})
     \times Y_{\ell m'}(\hat{r_j})]_{J=0} \}
\label{eq:surfwave}
\end{equation}

\noindent
in the $^{16}\mathrm{O}$ wave function.  However, values of $\alpha$
that had a significant (few percent) effect on $Z_h$ resulted in less
binding energy for $^{16}\mathrm{O}$ and hence are variationally excluded.

The $(e,e'p)$ cross sections are very sensitive to the $Z$ and the radius of
the
quasi-hole wave function. In Fig. 3 for example the experimental momentum
distribution for the $^{16}\mathrm{O}(e,e'p)^{15}\mathrm{N}$ leading to the
$(3/2)^-$
state
of $^{15}\mathrm{N}$ at 6.32 MeV is shown. The proton is ejected
quasi-elastically
with 90 MeV of kinetic energy in the center-of-mass system and with its
momentum lying in the $\vec{q}$ direction, i.e. in the so-called parallel
kinematics \cite{Leuschner}. By varying $q$ itself, it is possible to
extract the distribution of the
missing momentum $p_m$ (which is related to the momentum of the bound
nucleon) keeping the ejectile energy fixed, i.e. keeping the final state
interactions fixed.

The theoretical reduced cross sections obtained with
$\sqrt{0.86} \, \psi_h$ and $\psi_{WS}$ are shown with the dashed and solid
lines, respectively. The $\psi_{WS}$ gives a much better description of
data while $\sqrt{0.86} \, \psi_h$ overestimates them at
$p_m<200$ MeV/c. For both cases the ejected
proton wave
function $f_P \chi$ is given by the optical potential of Ref.
\cite{Schwandt} and the Coulomb distortion of electron waves has been taken
into account through the effective momentum approximation \cite{GP}.

\section{WAVE FUNCTION OF THE EJECTED PROTON}

In this section we study the effect of three possible improvements in
$\psi_p(x=\vec{r},\vec{\sigma},\vec{\tau}\, )$, the wave function of the
ejected
proton.  In the one-body representation it can be calculated, like the
$\psi_h(x)$ of Eq. (\ref{eq:specamp}), from the overlap

\begin{equation}
\psi_p(x) = \langle\Psi_h|a(x)|\Psi_f\rangle,
\label{eq:scattamp}
\end{equation}

\noindent
where $|\Psi_f\rangle$ is the $A$-nucleon final state asymptotically
corresponding to
an outgoing proton of energy $E_p$ leaving the nucleus in the $(A-1)$-nucleon
state
$|\Psi_h\rangle$.  In the analysis of NIKHEF data the $\psi_p(x)$ is
approximated by $f_P (r) \chi(x)$ with a Perey factor consistent with
the adopted phenomenological optical potential.

In mean-field (MF) approximation the outgoing distorted wave may be
calculated from
either a local energy-dependent, or a non-local momentum-dependent optical
potential.  The waves obtained from these two, respectively denoted by
$\chi(x)$ and $\chi'(x)$, are approximately related by

\begin{equation}
\chi'(x) = \sqrt{{m^*(\vec{r}\, ) \over m}}  \mbox{ } \chi(x) ,
\label{eq:chi}
\end{equation}

\noindent
where $m^*(\vec{r}\, )$ is the effective mass of the outgoing proton at
$\vec{r}$.
Outside the nucleus $m^* = m$, $\chi' = \chi$, and hence both give the same
nucleon-nucleus scattering observables.  However, inside the nucleus, where
$m^*(\vec{r}\, ) \neq m$, only the $\chi'(x)$ conserves the proton flux.
Therefore
$\chi'(x)$ should be used to calculate the transition matrix element.

The problem with using $\chi$, and the underlying physics of Eq.
(\ref{eq:chi}), is most
easily seen by considering a beam of nucleons impinging on nuclear matter
occupying
the $z>0$ half of space.  As usual we have incident, reflected and transmitted
waves denoted by $A \mathrm{e}^{\mathrm{i}kz}$, $B \mathrm{e}^{-\mathrm{i}kz}$
and $C \mathrm{e}^{\mathrm{i}k'z}$ respectively, where $k$ and
$k'$ are the momenta of the incident nucleons outside and inside nuclear
matter.  The flux conservation is given by

\begin{equation}
|C|^2 = (|A|^2-|B|^2) {k \over k'} \mbox{ } {m^*(k') \over m},
\label{eq:flux}
\end{equation}

\noindent
where $m^*(k')$ is the effective mass in nuclear matter.  When an
energy-dependent
local potential is used to describe the nucleon beam one obtains the
incorrect flux
conservation corresponding to $m^*(k') = m$ inside matter.

The effective mass $m^*(r)$, at energy $E_p$, is given by the well
known
equation \cite{Mahaux}

\begin{equation}
{m^*(r) \over m} = 1 - {\partial V(E,r) \over \partial
E}\mbox{ }\Big \vert_{E_{p}} \equiv f^2_P(r) ,
\label{eq:mstar}
\end{equation}

\noindent
which shows that the Perey factor is responsible for the flux conservation.
When the $V(E,r)$ is linear in $E$ over the entire range
$E_h \leq E \leq E_p$ one recovers the approximation for $\chi'(x)$ adopted in
the analysis of NIKHEF data.

In infinite nuclear matter the effective mass depends upon the density $\rho$
of
matter and the energy of the proton, or equivalently the momentum
$k(E,\rho)$.  The
function $m^*(k[E,\rho],\rho)$ has been recently studied using realistic
nuclear
forces \cite{R.B.W.88,V.R.P.}.  Using the local-density approximation (LDA)

\begin{equation}
m^*(\vec{r}\, ) = m^*(k(E_p,\rho(\vec{r}\, )), \rho (\vec{r}\, )),
\label{eq:mlda}
\end{equation}

\noindent
where $\rho(\vec{r}\, )$ is the density distribution of $^{16}\mathrm{O}$,
we can
replace the
Perey factor by $\sqrt{m^*(\vec{r}\, )/m}$ in the $\psi_p(x)$.  This change
has very
little effect on the calculated cross-section in the NIKHEF kinematics as
shown in
Fig.~4.

In the studies of the quasi-hole orbitals in helium liquid drops
\cite{Lewart}, it
was found that they could be related to MF orbitals with the LDA

\begin{equation}
\psi_h^{LDA}(x) = \sqrt {Z[\rho(\vec{r}\, )]} \mbox{ } \phi_h^{MF}(x) \approx
\psi_h(x) .
\label{eq:psilda}
\end{equation}

\noindent
The $\phi^{MF}(x)$ have a unit norm and are chosen so that

\begin{equation}
\sum_{\mathrm{occupied} \mbox{ } i} |\phi_i^{MF}(x)|^2 = \rho(\vec{r}\, ).
\label{eq:rho}
\end{equation}

\noindent
Interestingly the $f_P(r) \phi_{WS}(x)$ used to fit the NIKHEF data are
very similar to the $p$-wave $\phi^{MF}(x)$ required to reproduce the
$\rho$
of $^{16}\mathrm{O}$.  The $Z(\rho)$ in Eq. (\ref{eq:psilda}) is
interpreted as the
renormalization constant in matter at density $\rho$.  The $\psi_h^{LDA}$
obtained
with the linear expression

\begin{equation}
\sqrt{Z(\rho)} = 1 - \left(1 - \sqrt{Z_o}\right) \mbox{ } {\rho \over
\rho_o},
\label{eq:zetarho}
\end{equation}

\noindent
where $\rho_o$ is the equilibrium density $(0.16$ fm$^{-3})$ of matter and
$Z_o =
Z(\rho_o)$, is very close to the $\psi_h(x)$ for $Z_o = 0.64$, as shown in
Fig. 2.
This value of $Z_o$ is smaller than the $Z_o = 0.71$ estimated from detailed
calculations \cite{F.P.84} with the Urbana model of $\mathrm{v}_{NN}$.
However, most of
the difference could be because the Argonne model of $\mathrm{v}_{NN}$ used
in the present
work has a stronger tensor force.  The $\sqrt{Z(\rho)}$ takes into
account the
reduction of the overlap wave function (Eqs. \ref{eq:varqh},
\ref{eq:scattamp}) from its MF value due to short-range correlations.

Assuming that $\chi(x)$ is the MF wave function of the ejected proton, we
obtain an improved approximation

\begin{equation}
\psi_p(x) = \sqrt{(m^*(\vec{r}\, )/m) \  Z[\rho(\vec{r}\, )]} \mbox{ } \chi(x)
\label{eq:chizeta}
\end{equation}

\noindent
with the LDA of Eq. (\ref{eq:zetarho}).  The $(e,e'p)$ cross sections obtained
with this $\psi_p(x)$ are
smaller by $\sim 10\%$ than those obtained with $f_P(r) \chi(x)$, as shown in
Fig.~4. Naturally, if this $\psi_p$ is used to analyze the NIKHEF data the
extracted values of the spectroscopic factors for the states in
$^{15}\mathrm{N}$ will
be
larger by $\sim 10\%$ than those given in Ref.~\cite{Leuschner}.

The third improvement in $\psi_p$ is meant to take into account the difference
between the imaginary potential $W(\vec{r}\, )$ seen in nucleon-nucleus
scattering and
in
the $(e,e'p)$ reaction.  The $W(\vec{r}\, )$ in nucleon-nucleus scattering can
be
regarded as

\begin{equation}
W(\vec{r}\, ) = {1 \over 2} \sigma_{eff}(\vec{r}\, ) v(\vec{r}\, )
\rho(\vec{r}\, ),
\label{eq:imopt}
\end{equation}

\noindent
where $\sigma_{eff}(\vec{r}\, )$ is the effective $NN$ cross section, which can
have
spatial dependence due to density-dependent effects such as Pauli blocking,

\begin{equation}
v(\vec{r}\, ) = {\hbar k(\rho) \over m^*(k(\rho),\rho(\vec{r}\, ))}
\label{eq:vel}
\end{equation}

\noindent
is the local velocity, and the local momentum $k$ is given by

\begin{equation}
{\hbar^2k^2 \over 2m} = E_p-V(E_p,r).
\label{eq:ropt}
\end{equation}

In an $(e,e'p)$ reaction, let the ejected proton be struck at position
$\vec{r}_o$ and time $t_o$.  The distribution of nucleons of the residual
nucleus
at time $t_o$
is then given by the two-nucleon density $\rho_2(\vec{r}_o,\vec{r}\, )$.
Assuming that
it does not change significantly in the time taken by the ejected nucleon
to come out
of the nucleus, the imaginary potential $W'(\vec{r}\, )$ seen by the ejected
proton
is given by

\begin{equation}
W'(\vec{r}\, ) = {1 \over 2}
\sigma_{eff}(\vec{r}\, ) v(\vec{r}\, ) \rho_2(\vec{r}_o,\vec{r}\, )
   \equiv {1 \over 2}
\sigma_{eff}(\vec{r}\, ) v(\vec{r}\, ) \rho(\vec{r}\, ) g(\vec{r}_o,\vec{r}\,
),
\label{eq:imcopt}
\end{equation}

\noindent
where $g(\vec{r}_o,\vec{r}\, )$ is the pair distribution function.  In
practice
we must differentiate between the $pp$ and $pn$ distribution functions as
discussed in
\cite{V.R.P.}. However, this difference is suppressed here for brevity.
Due to the
repulsive core in the $NN$ interaction and the Pauli exclusion,
$g(\vec{r}_o,\vec{r}\, )<1$ at small $|\vec{r}_o-\vec{r}\, |$.  Therefore,
transparencies
calculated from $W'(\vec{r}\, )$ are larger than those from $W(\vec{r}\, )$
\cite{V.R.P.} in agreement with the data \cite{Garino}.

Let the $z$-axis be in the direction of the ejected proton and $\vec{r} =
(\vec{r}_{\perp},z)$.  In the Glauber approximation the damping of the
ejected proton
wave, emerging from $\vec{r}_o$, is given by

\begin{eqnarray}
D(\vec{r}_o,r\rightarrow\infty) &= \exp \left\{ - {1 \over 2}
{\displaystyle \int^\infty_{z_o}dz'\  \rho(\vec{r}\, ')\
\sigma_{eff}(\vec{r}\, ')}
\right\} \nonumber
\\
  &= \exp \left\{ {\displaystyle - \int^{\infty}_{z_o}dz'\  {W(\vec{r}\, ')
\over
v(\vec{r}\, ')}  } \right\},
\label{eq:damp}
\end{eqnarray}

\noindent
when the nucleon-nucleus optical potential is used.  However, the correct
damping obtained from $W'(\vec{r}\, )$ is

\begin{equation}
D'(\vec{r}_o,r\rightarrow\infty) = \exp\left\{-
\int^{\infty}_{z_o}dz'\  {W(\vec{r}\, ') \over
v(\vec{r}\, ')} \  g(\vec{r}_o,\vec{r}\, ')\right\}.
\label{eq:dampc}
\end{equation}

\noindent
The effect of the increase in the transparency for $(e,e'p)$ reaction can be
easily
incorporated in the calculation of the cross section by modifying the
distorted
wave $\chi(x)$:

\begin{equation}
\tilde{\chi}(x) = \lambda(\vec{r}_{\perp},z)\chi(x)
\label{eq:chilam}
\end{equation}

\begin{equation}
\lambda(\vec{r}_{\perp},z) = \exp\left\{+ \int^{\infty}_{z_o}dz'\  {W(\vec{r}\,
')
\over v(\vec{r}\, ')} \  (1-g(\vec{r},\vec{r}\, '))\right\}.
\label{eq:lambda}
\end{equation}

The $\psi_p(x)$ including the effects of increased transparency and the
final state
correlations, i.e. all three of the above improvements,
is given by

\begin{equation}
\psi_p(x) = \left[ Z(\vec{r}\, ) m^*(\vec{r}\, )/m \right]^{1/2} \
\lambda(\vec{r}\, ) \
\chi(x) .
\label{eq:chicor}
\end{equation}

\noindent
The cross sections obtained with it,
shown in
Fig. 4, are $\sim 3\%$ smaller than those obtained with $f_P(r) \chi(x)$.
It thus appears
that improvements in the wave function of the struck proton used in the
analysis of
the $(e,e'p)$ data may increase the extracted values of spectroscopic factors
\cite{Leuschner} by a few percent.

\section{CONCLUSIONS}

Overlaps of variational
nuclear wave functions that include realistic non-central correlations
appear to give a reasonable description of the quasi-hole wave function in
the interior of the nucleus. However, the resulting $\psi_h$
is too large in the surface. For a light nucleus like $^{16}\mathrm{O}$, about
half of the reduction of the $Z$ (the spectroscopic factor) from unity
comes from center-of-mass effects.
Approximate inclusion of correlation corrections
to the optical model wave functions traditionally used to analyze $(e,e'p)$
reactions appears to increase the extracted spectroscopic factors
by a few percent.  Nevertheless, a calculation using presently
available variational wave functions for $^{16}\mathrm{O}$ and
$^{15}\mathrm{N}$ and
these approximate corrections substantially overpredicts the observed cross
sections.

\section{ACKNOWLEDGEMENTS}

We thank professor L. Lapik\'as for providing us with the
$^{16}\mathrm{O}(e,e'p)$
data prior to publication.

This work was supported in part by the U.S. Department of Energy, Nuclear
Physics Division, under Contract No. W-31-109-ENG-38, and by the U.S.
National Science Foundation, Grant No. PHY 89-21025.

The calculations
were made possible by generous friendly-user access both to the IBM SP1
computer at the Mathematics and Computer Science Division (Argonne
National Laboratory) and to the HP 9000 cluster at the INFN - Sezione di Pavia
(Pavia, Italy), and by grants of computer time at the National Energy
Research Supercomputer Center (Livermore, California) and the National
Center for Supercomputing Applications (Urbana, Illinois).

\eject

\centerline{CAPTIONS}
\vskip .5 truecm

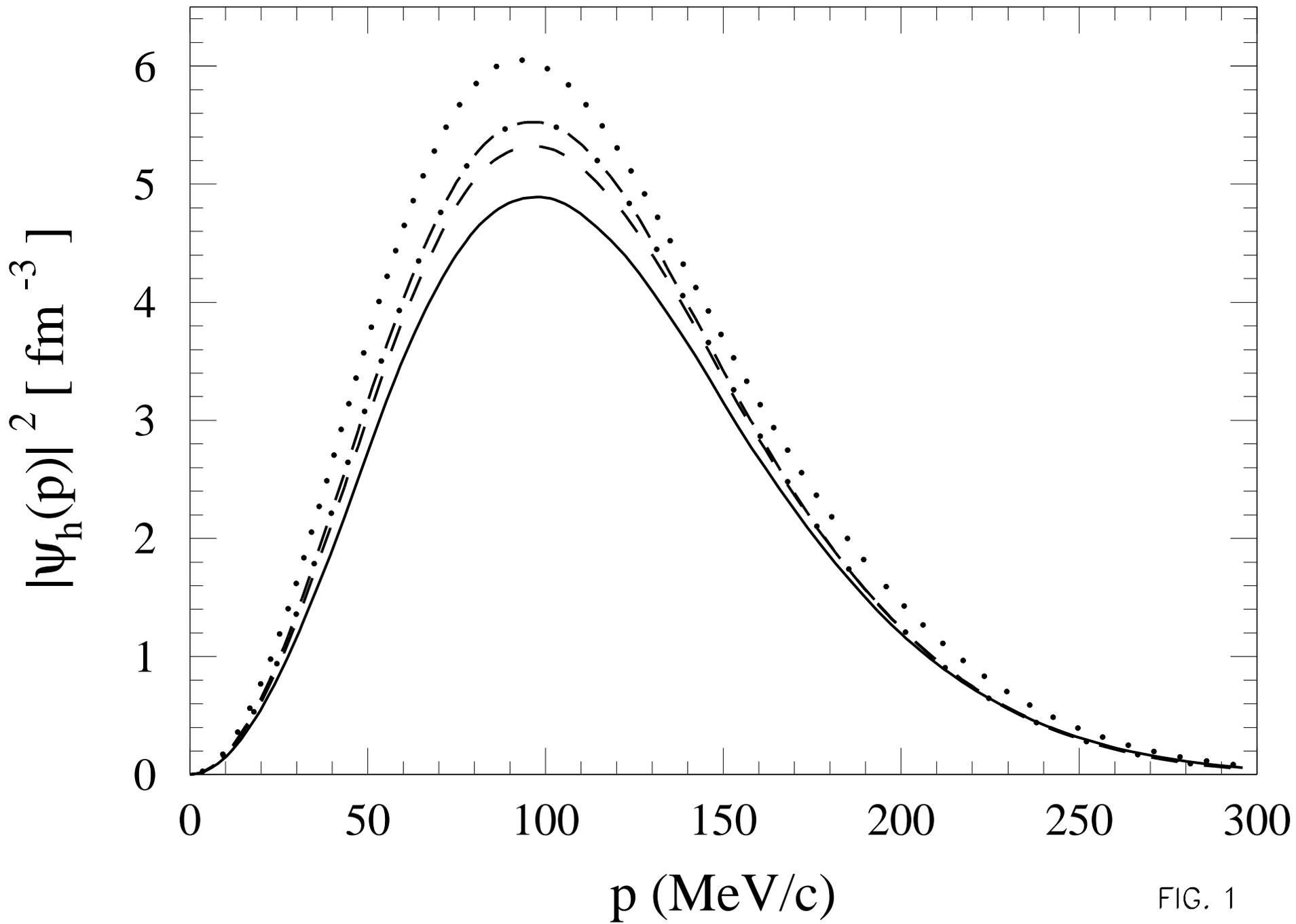
FIG. 1.  Various approximations to the $J=3/2$ quasi-hole $\vert \psi_h
(p) \vert^2$. The dotted line is the simple $p$-wave single-particle wave
function normalized to unity. The dashed-dot line includes also the
center-of-mass corrections. The dashed line shows the further effect of
central
correlations, while the solid line is for the full wave function including the
non-central correlations.

\vskip 10pt

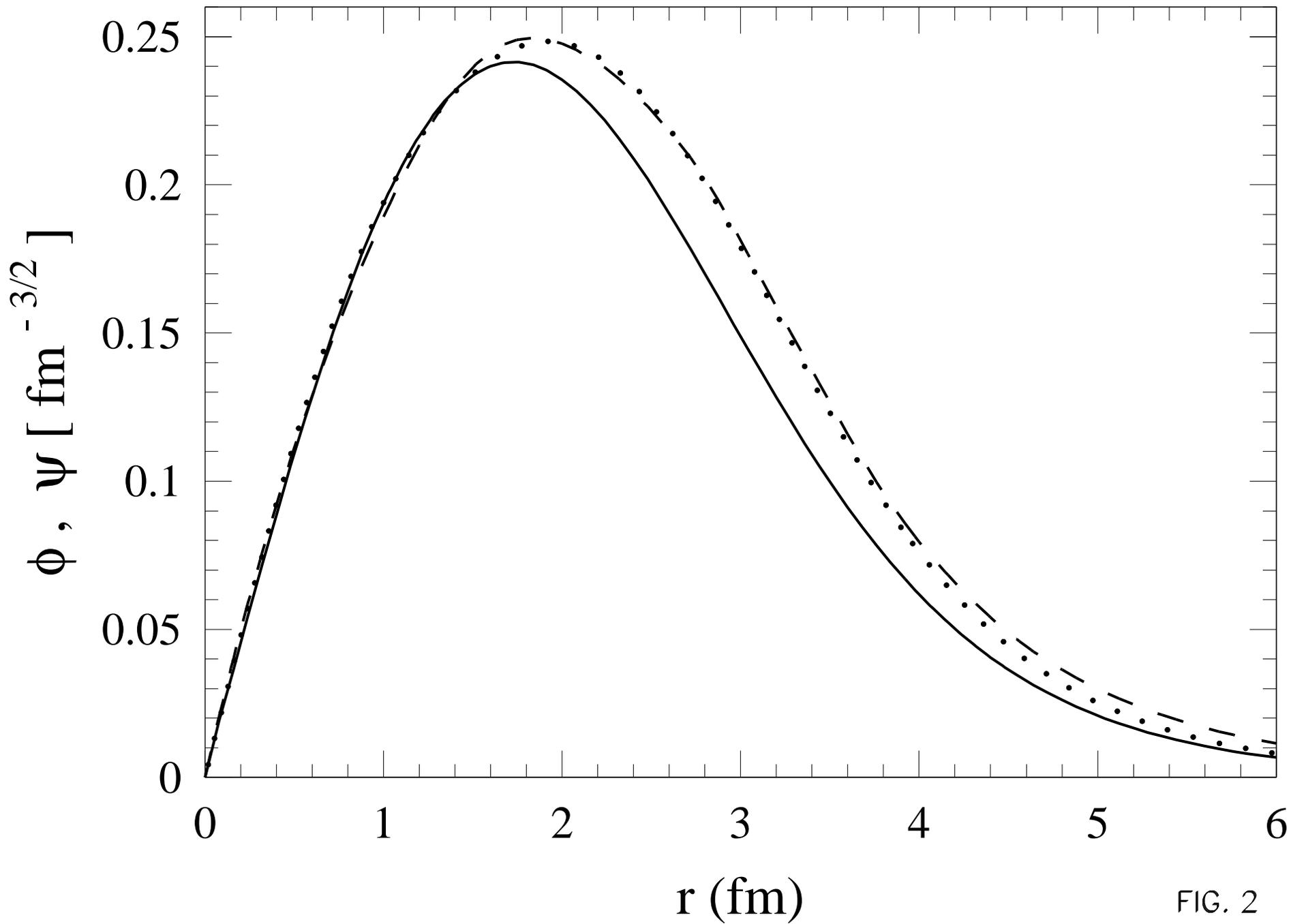
FIG. 2.  Different models for the single--particle bound state wave
function of the $p_{3/2}$ hole in $^{16}\mathrm{O}$. The dashed line is for the
quasi--hole $\sqrt{0.86} \  \psi_h$, the solid line is for the effective
Woods-Saxon $\psi_{WS}$, the dotted line is for the LDA
$\sqrt{0.86} \  \psi_h^{LDA}$ [see text and Eqs. (\ref{eq:psilda}),
(\ref{eq:zetarho})].

\vskip 10pt

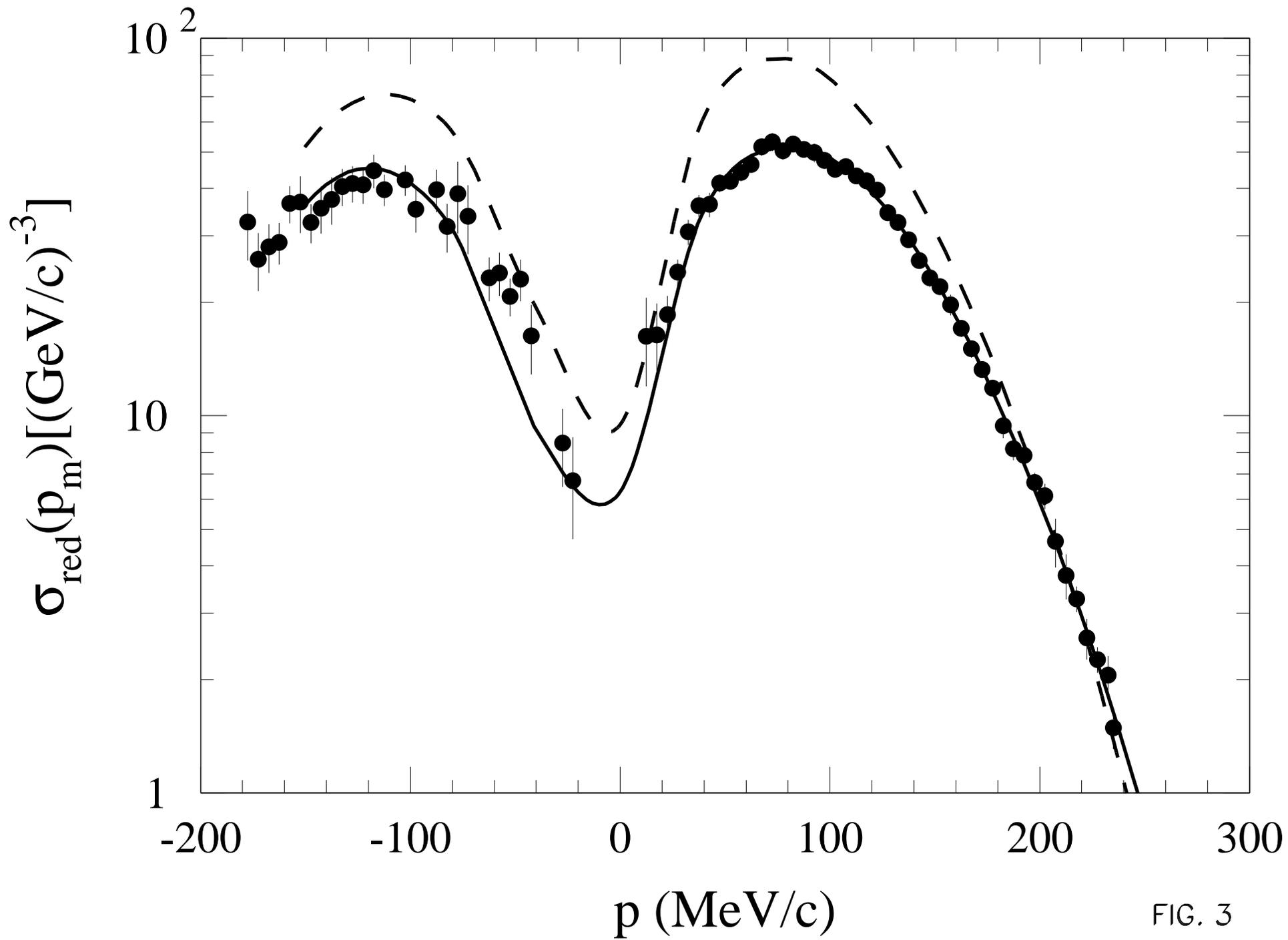
FIG. 3.  Theoretical reduced cross sections for the
$^{16}\mathrm{O}(e,e'p)^{15}\mathrm{N}$ reaction leading to the  $3/2^-$
state at 6.32
MeV.
The proton is ejected with 90 MeV of center--of--mass energy in quasi--elastic
parallel kinematic conditions. The solid line adopts the effective Woods-Saxon
bound state $\psi_{WS}$ while the dashed line uses the quasi-hole $\psi_h$ (see
text).
The scattering state is from the optical potential of Ref. \cite{Schwandt} and
is
multiplied by the proper Perey factor. The data are from Ref.
\cite{Leuschner}.

\vskip 10pt

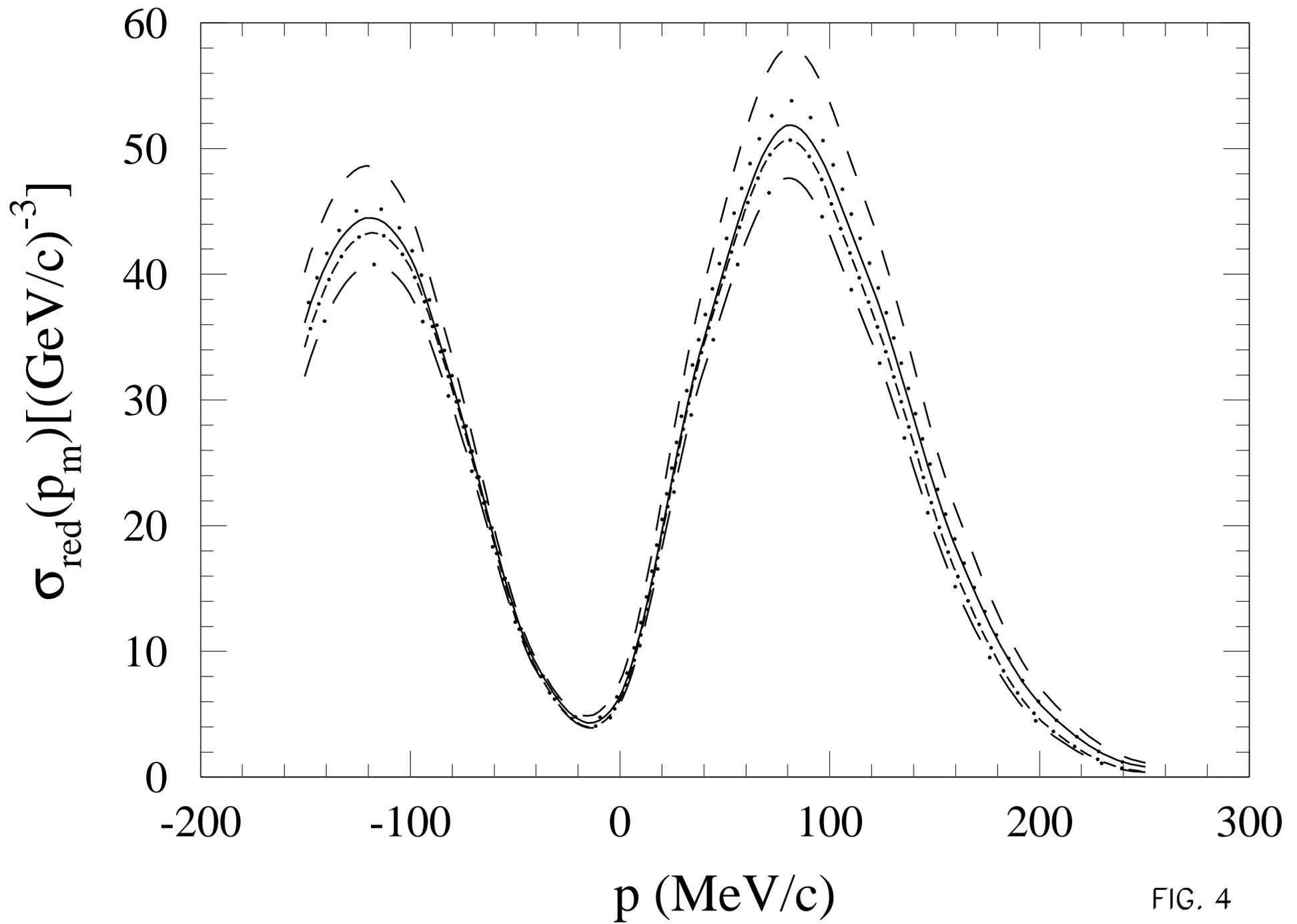
FIG. 4.  Theoretical reduced cross sections for the same reaction in the
same
kinematics as in Fig. 3. The solid line is the same as in Fig. 3. The other
curves
show the result for different modifications of the final scattering state. The
dashed line neglects the Perey factor. The dotted line substitutes the
equivalent effective mass correction for it. The long dash-dot line
further adds the
LDA correction [Eq. (\ref{eq:chizeta})] for short-range correlations. Finally,
the
short dash-dot line further adds the correction for the
correlation-hole on the final state [Eq. (\ref{eq:chicor})].

\end{document}